\documentclass{ws-procs975x65}

\begin{document}

\title{Gamma-Ray Burst Progenitors Confront Observations}

\author{Davide Lazzati}

\address{Institute of Astronomy, University of Cambridge\\
Madingley Road, CB3 0HA, Cambridge\\
United Kingdom}

\maketitle

\abstracts{The discovery of a supernova emerging at late times in the
afterglow of GRB~030329 has apparently settled the issue on the nature
of the progenitor of gamma-ray bursts. We now know that at least a
fraction of cosmological GRBs are associated with the death of massive
stars, and that the two explosions are most likely simultaneous. Even
though the association was already suggested for GRB~980425, the
peculiarity of that burst did not allow to extend the association to
all GRBs. The issue is now to understand whether GRB~030329 is a
``standard burst'' or not. I will discuss some peculiarities of
GRB~030329 and its afterglow lightcurve showing how, rather than a
classical cosmological GRB, it looks more like a transition object
linking weak events like GRB~980425 to the classical long duration
GRBs. I will also discuss the problems faced by the Hypernova scenario
to account for the X-ray features detected in several GRBs and their
afterglows.}

\section{Introduction}

The progenitor of Gamma-Ray Bursts (GRBs) - i.e. the astronomical
object that is associated to the energy release powering the GRB
emission, and that probably disappears in the process - has been a
major unknown since their discovery in the late sixties\cite{kle}.
Since the discovery of afterglow emission\cite{cos,van}, that made
possible their precise localisation in the sky, several circumstantial
pieces of evidence have been collected, linking the burst emission
with massive star formation phenomena\cite{kul,fru,hol,blo,laz1}. Even
though such studies pointed toward an association of GRB explosions
with the death of massive stars, the issue was far from being solved,
the main worry being that multi-wavelength modelling of afterglows
yielded typically a uniform ambient medium\cite{pan}, contrary to the
stratified wind expectations for the massive star
association\cite{che}.

The exact association of the burst with the star death was therefore
put under debate. The simplest scenario (Hypernova or Collapsar) would
call for a single explosion, in which the GRB would be produced by a
relativistic jet propagating along the rotational axis of a fast
spinning star, which explodes as a more normal supernova along the
equator\cite{mac}. Alternatively, the two explosions (SN and GRB)
could be separated by a relatively short interval of time (Supranova),
during which a meta-stable compact object is left, whose eventual
collapse cause the GRB explosion\cite{vie}. Finally, new population
synthesis studies showed that neutron star binary systems may be short
lived, allowing for the GRB explosion within the host galaxy\cite{per}
even in the classical binary merger scenario\cite{eic}.

The association of the peculiar GRB~980425 with SN1998bw\cite{gal}
and, more recently, that of GRB~030329 with SN2003dh\cite{sta,hjo} and
of GRB021211 with SN2002lt\cite{del} has given strong support to the
idea of a single explosion, which simultaneously generates a GRB and a
particularly energetic SN explosion. There is however a number of
observations that seem to point to a more complex association. The
issue is therefore to understand whether GRB~980425 and GRB~030329 are
``classical GRBs'' and to which extent all the observations can be
incorporated in a single coherent picture.

In this review I will critically discuss some aspects of this problem,
underlying possible peculiarities of the prompt and afterglow emission
of the two bursts robustly associated to SN explosions. I will then
discuss progenitor indications from X-ray spectroscopy of the prompt
and afterglow emission, discussing the problems that arise when we
attempt to include these observations in a simple Hypernova scenario

\section{Collapsars \& Hypernov\ae}

The observation of non-thermal high energy power-law tails in GRB
spectra, well above the pair production threshold
$h\nu=m_ec^2=511$~keV allows us to draw to fundamental
conclusions. First, the photon producing medium must be outflowing at
hyper-relativistic speed, with $\Gamma\gtrsim100$, to avoid the
absorption of photons above threshold. Second, to preserve the
non-thermal character of the spectrum, the moving plasma must contain
a small but sizable amount of baryons, of the order of
$M_0\sim10^{-4}M_\odot$. In other words, the baryon contamination of
the fireball, even if non negligible, must be extremely small. This
limit is particularly stringent for any model in which the
relativistic GRB jet has to find its way through a massive star.

A wide relativistic jet (where wide means such that
$\theta_j>\Gamma^{-1}$) that propagates through a medium collects all
the material ahead of itself since, due to the relativistic aberration
of trajectories, the stellar material cannot flow to the sides of the
jet\cite{maz}. For this reason, a relativistic jet would be engulfed
with the stellar material and effectively slowed down to
sub-relativistic speed. Even if the jet is generated as a relativistic
flow in the core of the star, it will drive a bow shock at its head,
which will advance at sub-relativistic speed pushing the dense stellar
matter on the sides. In this process part of the jet bulk energy is
randomised and lost from the relativistic flow. It may be recycled in
a second time, when the engine turns off, in the form of a delayed
wide-angle fireball component\cite{ram}.

\begin{figure}[t]
\centerline{\epsfxsize=4.1in\epsfbox{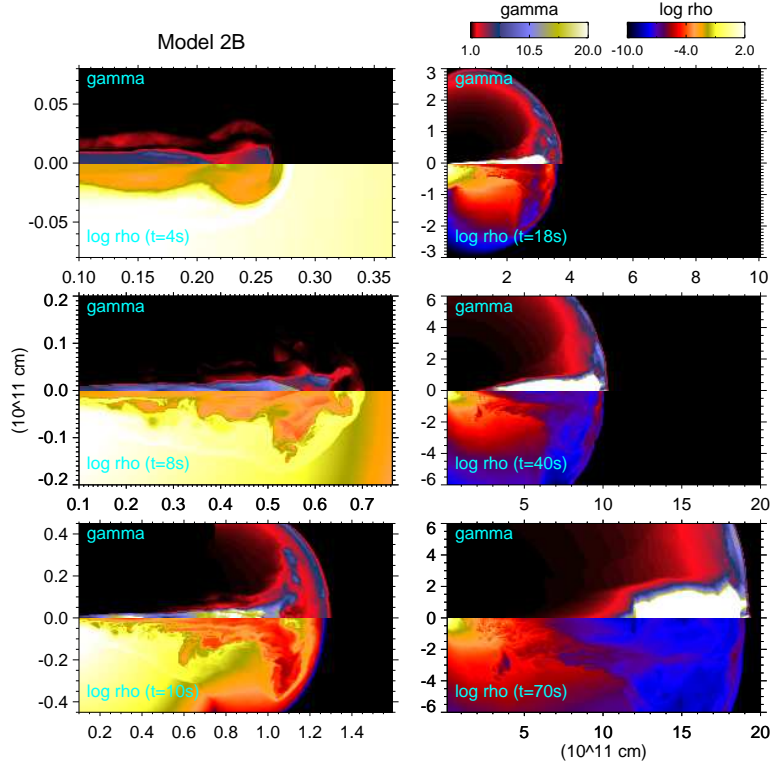}}
\caption{Results of relativistic simulations for the propagation of a
jet in a compact stellar core. From Zhang et al.\protect\cite{zha}. 
\label{fig:jet}}
\end{figure}

Numerically, the jet evolution has been studied extensively under
varied conditions, both to understand how it propagates into the
star\cite{mac,zha} and the effect on the stellar explosion\cite{kho}.
Figure~\ref{fig:jet} shows some of the more recent results by Zhang et
al.~\cite{zha}. In all these studies the jet is postulated at the core
of the star while, more recently, MHD simulations for the jet
formation have been performed\cite{pro}.

A first constraint on the properties of the progenitor star is its
radius. Being non-relativistic, the jet spends a time\footnote{Here
and in the following we adopt the notation $Q=10^x\,Q_x$ and use
c.g.s. units.}  $t\gtrsim{}R_\star/c\sim3\,R_{\{\star,11\}}/c$ to
propagate through the star mantle. During all this time, the inner
engine has to be on, to sustain the propagation of the jet head. This
time is lost, in the sense that the energy released is not seen in the
GRB emission. In order to avoid a fine tuning in the distribution of
the duration of long GRBs and to explain the ``standard GRB
energy''\cite{fra,pan}, we must conclude that the propagation time is
much shorter than the burst duration, and therefore
$R_\star<10^{11}$~cm\cite{mac}. The star has therefore lost all the
hydrogen and helium envelopes through a wind (Wolf Rayet phase) or the
interaction with a close companion in a binary system\cite{izz}. Since
a massive wind would probably also cause the loss of angular momentum
(another fundamental ingredient in the recipe for successful jet
propagation), the latter possibility seems favoured.

In terms of populations, there are many more type Ibc SNe than GRB
explosions, and therefore the loss of the envelope seem not to be the
sufficient ingredient for a successful jet propagation, confirming the
idea that an additional condition must be fulfilled. Interestingly, to
date, the only robust associations with GRBs are for type Ic SNe,
these without trace of either hydrogen and helium in their
spectra\cite{gal,sta,hjo}.

\section{Supranov\ae}

Supranov\ae~where proposed as possible progenitors for GRB explosions
in order to solve the problem of baryon contamination without losing
the association of GRBs with the death of a massive star. The main
characteristic feature of a supranova explosion is that is made by two
explosions. At the end of its life, a massive star explodes as a SN,
leaving behind a compact unstable object. This, in the most popular
version of the model, is a massive neutron star (NS), too massive to
be stable and avoid the implosion into a black hole. Most NS equations
of state, however, allow for meta-stable super-massive objects if the
NS is fastly spinning, so that the centrifugal support provides the
extra pressure is required to keep the configuration stable\cite{vie}.
If the newly born NS has a sizable magnetic field which is not
perfectly aligned with its rotation axis, the NS will lose energy as a
pulsar, at the expenses of rotational energy. As the NS slows down its
rotation, the condition for stability is lost and the system collapse
into a black hole. This collapse is supposed to power the GRB.

Vietri and Stella\cite{vie} computed the lifetime of the system to be:
\begin{equation}
t_{\rm sd}\equiv{{J}\over{\dot J}} = 10 \, {{j}\over{0.6}}
\left({{M}\over{3\,M_\odot}}\right)^2 \,
\left({{15\,{\rm km}}\over{R_{\rm NS}}}\right)^6 \,
\left({{10^4\,s^{-1}}\over{\omega}}\right)^4 \,
\left({{10^{12}\,{\rm G}}\over{B}}\right)^2 \,\,\,\,{\rm yr}
\end{equation}
and therefore the GRB explosion site should, in this model, be
surrounded by a relatively young SNR of radius:
\begin{equation}
R_{\rm SNR} \sim 3\times10^{17} \, {{j}\over{0.6}}
\left({{M}\over{3\,M_\odot}}\right)^2 \,
\left({{15\,{\rm km}}\over{R_{\rm NS}}}\right)^6 \,
\left({{10^4\,s^{-1}}\over{\omega}}\right)^4 \,
\left({{10^{12}\,{\rm G}}\over{B}}\right)^2 \,
\left({{v_{\rm SNR}}\over{10^4\,{\rm km/s}}}\right) \,\,\,{\rm cm}
\end{equation}
the result is strongly dependent on initial conditions, and therefore
the model predicts a range of time delays (and consequent SNR
radii). The presence of the nearby SNR, as we shall see in the
following, is particularly important to explain some of the observed
X-ray features. The explosion of the GRB inside a magnetised cavity
may as well ameliorate the problem of magnetic field generation and
explain the uniform ambient media derived from afterglow
modelling\cite{koe}.

\section{Prompt emission}

The association of GRB~030329 with the supernova SN2003dh bears much
more important consequences than that of GRB~980425 with
SN1998bw. While GRB~980425 is a peculiar GRB, with an energy budget
which is largely smaller than that of classical GRBs, GRB~030329 is,
at face value, a typical cosmological burst. In this section and in
the following I discuss the properties of GRB~030329 and compare them
to those of high redshift GRBs.

\subsection{Standard Energy}

\begin{figure}[t]
\centerline{\epsfxsize=4.1in\epsfbox{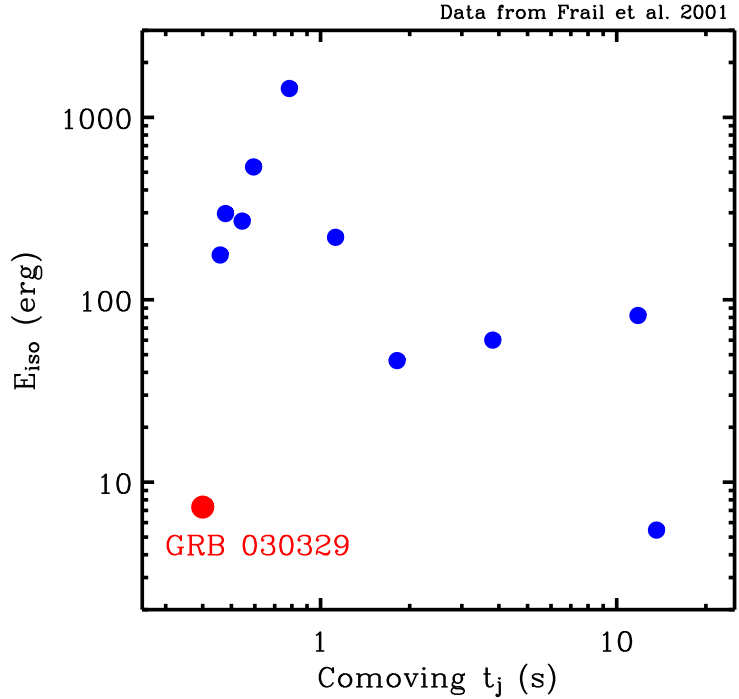}}
\caption{Isotropic equivalent energy release in $\gamma$-rays vs. jet
break time for a sample of ``classical'' GRBs and GRB~030329, which is
under-energetic in this plane.\label{fig:frail}}
\end{figure}

One of the most important results in GRB studies of the lat few years
is the realisation that the total energy budget of the event, once
corrected for the opening angle of the jet\footnote{In the framework of
the universal structured jet the correction is performed on the
viewing angle rather than on the opening angle of the jet\cite{ros}.}
is fairly typical, amounting to $E\sim10^{51}$~erg\cite{fra,pan}.

This value is certainly not attained by GRB~980425 which, even for a
spherical explosion, had a total energy release in ultra-relativistic
material $E\lesssim10^{48}$~erg. The case of GRB~030329 is more
complicated. The isotropic energy release of GRB~030329 is fairly
typical for a dim burst, but the afterglow light curve has a clear
break at early times $t_b\sim0.4~d$, which allows us to infer a small
opening angle. In Fig.~\ref{fig:frail} I show the correlation between
the isotropic equivalent energy and the break time for several
classical GRBs (which form a clear correlation) and GRB~030329. It is
clear that, given the break time, the energy is smaller by more than
an order of magnitude with respect to what predicted by the
correlation. 

\begin{figure}[t]
\centerline{\epsfxsize=4.1in\epsfbox{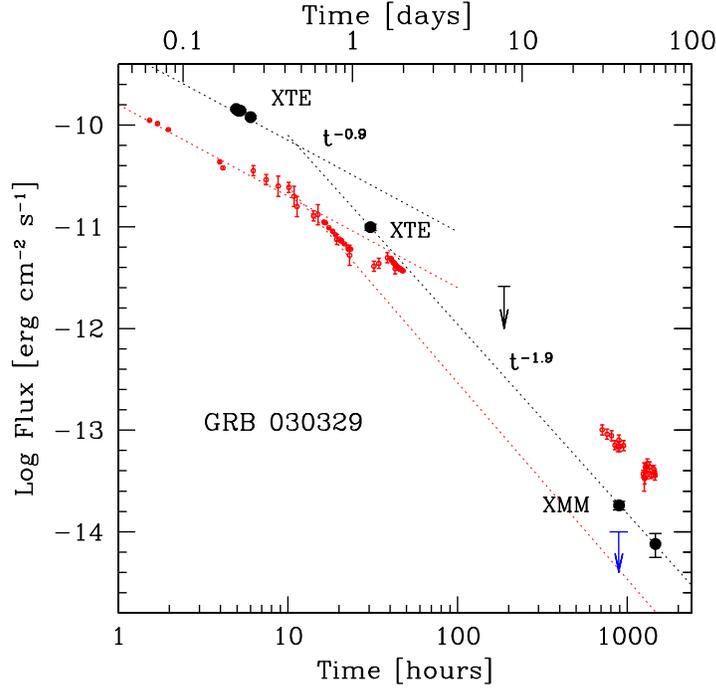}}
\caption{Optical and X-ray light curves of GRB~030329, showing that the
break in the optical is simultaneous with the break in the X-rays, as
predicted for a jet transition. From Tiengo et
al.\protect\cite{tie}.\label{fig:tiengo}}
\end{figure}

Given the complexity of the afterglow light curve of GRB~030329 (see
below) it is however worth asking ourselves whether the break is
really a jet break and not a spectral transition or merely a
fluctuation on top of a regular power-law decay. Fortunately several
X-ray observations, both from the Rossi-XTE and XMM, are available to
confirm that the break is achromatic\cite{tie}, as predicted for a jet
break. The break in GRB~030329 is actually the best example of an
achromatic break in an afterglow we have to date (see
Fig.~\ref{fig:tiengo}).

\subsection{Typical photon energy}

\begin{figure}[t]
\centerline{\epsfxsize=4.1in\epsfbox{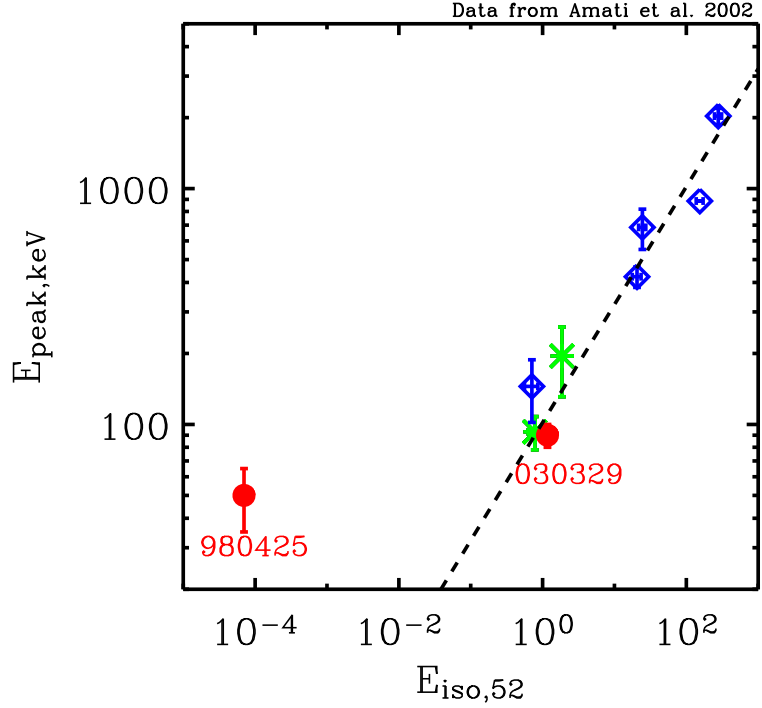}}
\caption{Isotropic equivalent energy versus peak photon frequency (in
$\nu\,F(\nu)$ representation). Diamonds show high redshift events, for
which no SN association is claimed, asterisks show intermediate
redshift events, for which the SN association is based on red bumps in
the late afterglow evolution. GRB~030329 and GRB~980428 are shown with
dots and indicated.\label{fig:amati}}
\end{figure}

Another interesting correlation that was discovered in classical long
GRBs is the correlation between the isotropic equivalent energy and
the peak of the $\nu\,F(\nu)$ spectrum\cite{llo,ama}. The correlation
is shown in Fig.~\ref{fig:amati}, where GRB~030329 and GRB~980425 are
indicated individually. While GRB~030329 fits the correlation,
fostering its appertaining to the classical GRB family, GRB~980428
stands out, underlying again its peculiarity.

\section{Afterglow}

\begin{figure}[t]
\centerline{\epsfxsize=4.1in\epsfbox{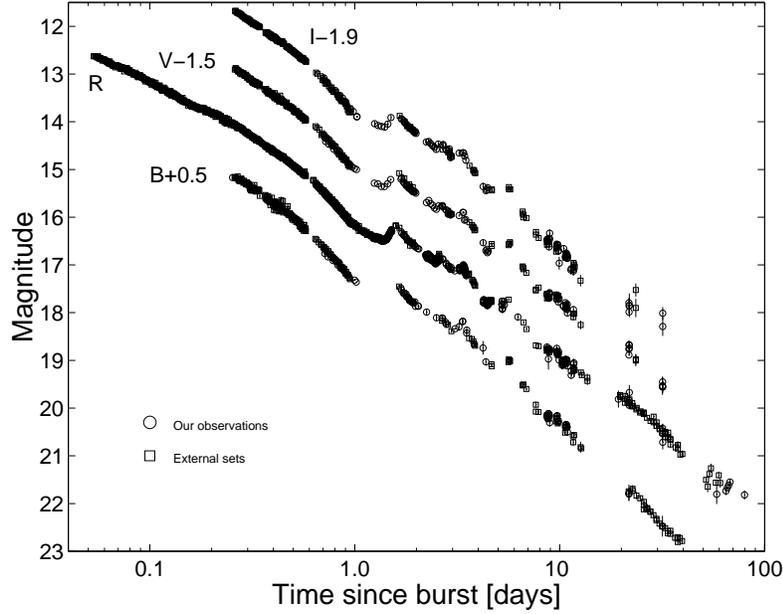}}
\caption{Optical BVRI light curve of GRB~030329. The amount of
photometric measurements performed on this event is unprecedented and
the light curve reveals a complex behaviour on top of the usual broken
power-law decay. From Lipkin et
al.\protect\cite{lip}.\label{fig:0329}}
\end{figure}

The phase in which GRB~030329 shows more peculiarities is the
afterglow. Whether this is due to an intrinsic difference with respect
to other GRBs or rather to a difference in the conditions of the
ambient medium is not clear and worth a deeper investigation. 

\subsection{Bumps and wiggles}

The multi-filter afterglow light curve of GRB~030329 is shown in
Fig.~\ref{fig:0329}. A first important thing to understand is whether
the complexity of the afterglow light curve is intrinsically larger
than in other GRBs or simply more apparent given the enormous
observational effort devoted to this particular afterglow\cite{lip}.
In comparison one can think to the case of GRB~020813, the smoothest
afterglow studied so far\cite{lau,gor}, for which 55 photometric
V-band observations were collected.

\begin{figure}[t]
\centerline{\epsfxsize=4.1in\epsfbox{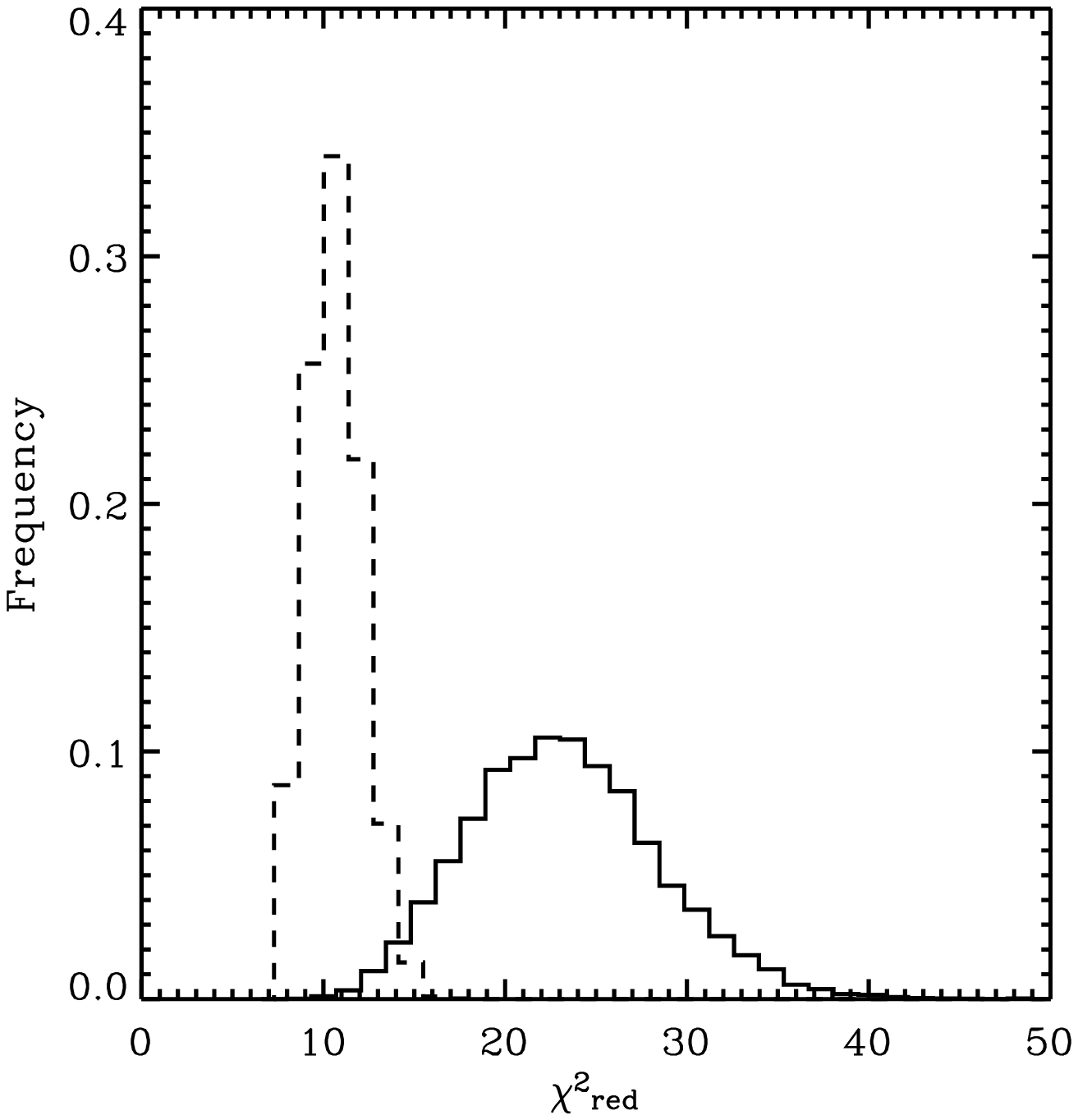}}
\caption{Distribution of $\chi^2$ values for two set of 10000
simulations each of reduced quality light curves drawn from real R-band
observations of GRB~030329 (see text for more details). The solid
histogram refers to simulations with 55 observational points without
reducing the accuracy of the measurement, while the dashed histogram
is for simulations where a minimum uncertainty of 0.05 magnitudes was
assumed for the photometry. \label{fig:chi}}
\end{figure}

To test the possibility that the light curve of GRB~020813 appears
smooth only due to a lack of coverage we have simulated 10000 fake
R-band light curves by randomly selecting 55 photometric measurements
from the $\sim1600$ from the light curve of GRB~030329 (data table
taken from Lipkin et al.\cite{lip}). For each simulated light curve we
have attempted to fit a smoothly broken power-law shape and estimated
the goodness of fit with the reduced $\chi^2$. The procedure was
repeated twice. First we assigned the real error bars to the
data. Subsequently the photometric accuracy was reduced by assigning a
minimum uncertainty of 0.05 magnitudes to each point. The distribution
of the resulting reduced $\chi^2$ values is reported in
Fig.~\ref{fig:chi}. For no single case out of 20000 total simulations
an acceptable $\chi^2$ was obtained, not even with the reduced
accuracy of the photometry. We are therefore able to conclude that the
afterglow of GRB~030329 represents a case of intrinsically high
variability, much larger than what observed in most cosmological
afterglows. To further deepen this analysis we reduced the number of
photometric observations. We find that only if the number of
photometric points is reduced to less than 10 the simulated
light curves can be effectively described as broken power-laws:
$\sim2\%$ of the light curves with 10 photometric points and
$\sim11\%$ of these with 5 yield $\chi^2_\nu\le2$.

An interesting issue is the origin of the observed afterglow
complexity. One important observation that helps reducing the
possibilities is the fast rise-time of the bumps. Any event (increase
or decrease of brightness) that takes place over the entire fireball
surface will be observed to smoothly affect the afterglow light curve
over a timescale $\delta\,t\sim{}T$, where $T$ is the moment in which
the deviation begins\cite{laz2}. Since the increase in brightness in
GRB~030329 is much faster it must be due to a local
phenomenon. Another clue comes from the fact that the fireball is
observed to monotonically brightens instead of showing random
increases and decreases in brightness. For these reasons the best
solution seems to be a late time input of energy from the inner
engine\cite{gra}, under the form of delayed shells. Since the extra
energy is input in the fireball after the jet break time, it does not
involve the entire fireball surface and can be characterised by a rise
time $\delta\,t<T$. A very important implication of this model is
that, given the prominence of the rebrightenings, the extra energy
input is larger by an order of magnitude than the original ejected
energy. If this extra energy is considered in Fig.~\ref{fig:frail},
the GRB~030329 point would move upward by more than a factor of ten,
making GRB~030329 consistent with the correlation.

The only problem for this interpretation is the polarization
curve\cite{gre}. If the extra energy input is always given to the same
part of the fireball, the polarization curve should show a marked
correlation between the rebrightenings and the polarization intensity
and position angle, something that is not observed in the
data. However, it is possible to argue that the refreshed shock will
not be completely uniform, giving rise to unpredictable fluctuations
in the polarization signal.

\subsection{Structure?}

An alternative interpretation for the complexity of the GRB~030329
afterglow light curve can be formulated based on the peculiarity of
its radio light curve. It shows a prominent break at
$t\sim10$~d\cite{ber}, which has all the properties of a second jet
break. Berger et al.\cite{ber} interpret it as due to a second
component in the fireball, with lower energy per unit solid angle and
Lorentz factor, but larger opening angle and larger total energy by a
factor 10. It turns out that a refined analysis of the optical
lightcurve requires as well a second break at $t\sim10$~d, as expected
for a jet break. The two components of the fireball may have been
simultaneous, in a structured jet scenario\cite{ros} or delayed in
time as well, the slower one being produced by the recycling of the
energy the jet wasted in the process of opening up a clean funnel in
the star\cite{ram}. Even though this interpretation explains the
overall shape of the light curve, it faces several problems to explain
the rapidity of fluctuations and the fact that not a single
rebrightening is observed. The extra ingredient of delayed energy
input is therefore still necessary. Also in this case, however, the
standard GRB energy of $\sim10^{51}$~erg would be recovered.

\begin{figure}[t]
\centerline{\epsfxsize=4.1in\epsfbox{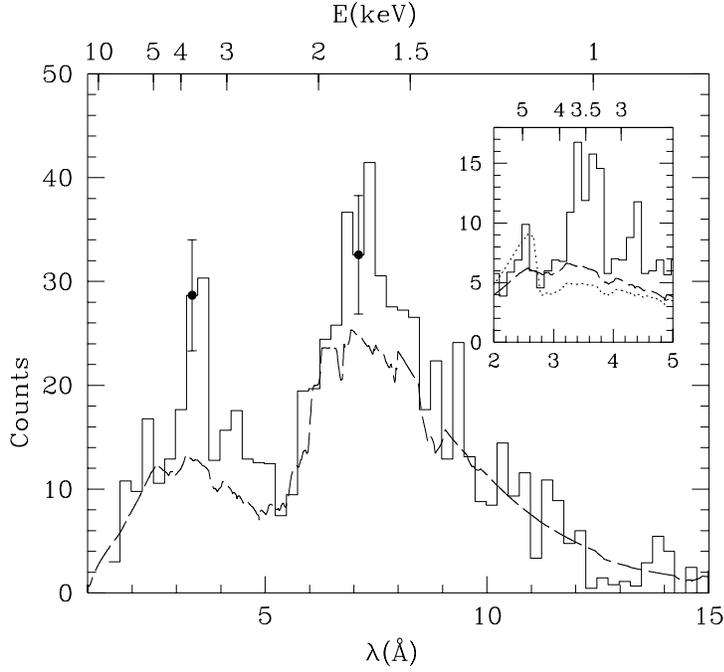}}
\caption{The Chandra grating spectrum of the early X-ray afterglow of
GRB~991216 showing evidence of a broadened X-ray emission line. From
Piro et al.\protect\cite{pir2}.
\label{fig:1216}}
\end{figure}

\section{X-ray features}

\begin{figure}[t]
\centerline{\epsfxsize=4.1in\epsfbox{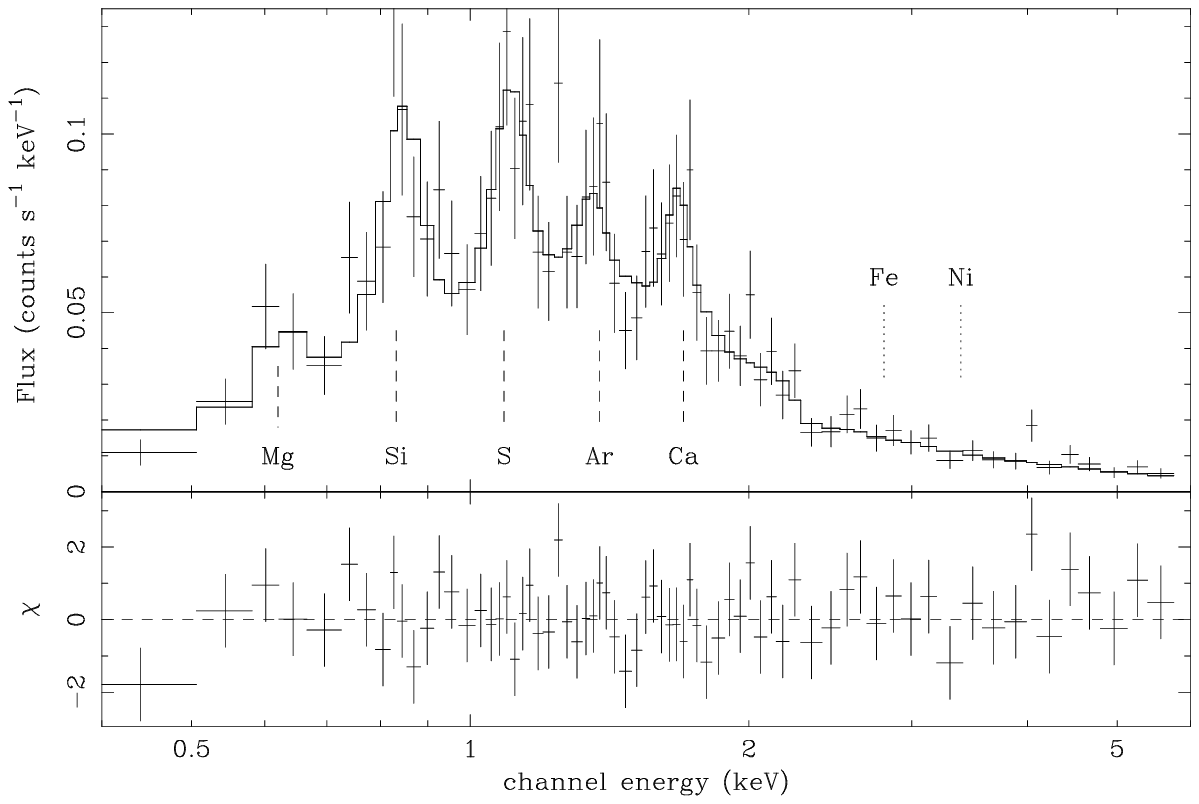}}
\caption{The XMM-Newton spectrum of the early X-ray afterglow of
GRB~030227 showing evidence of a complex of soft X-ray emission lines. From
Watson et al.\protect\cite{wat}.
\label{fig:0227}}
\end{figure}

\begin{figure}[t]
\centerline{\epsfxsize=4.1in\epsfbox{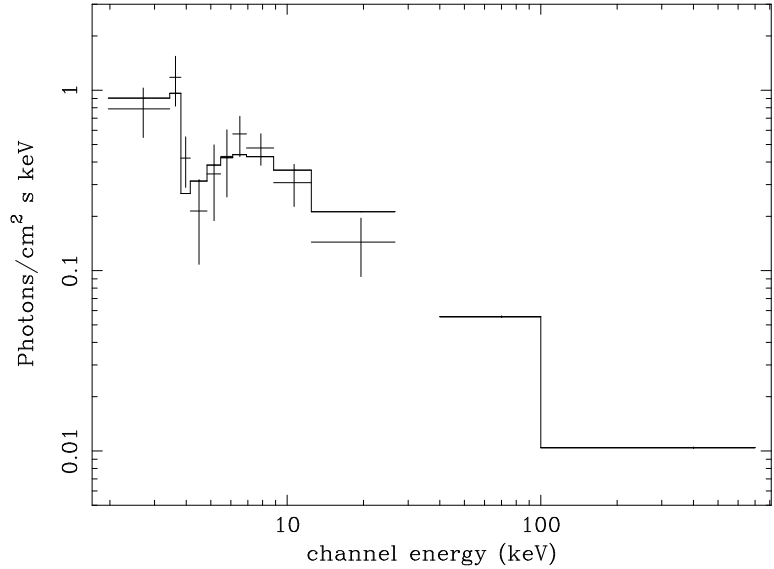}}
\caption{The BeppoSAX-WFC and GRBM combined spectrum of the early
X-ray prompt emission of GRB~990705 showing evidence of a deep
absorption feature. From Amati et al.\protect\cite{ama2}.
\label{fig:0705}}
\end{figure}

The most serious challenge to extending the hypernova origin of
GRB~030329 to all cosmological GRBs is represented by the possible
presence of X-ray narrow features, both in absorption and emission, in
the prompt and early afterglow emission of several
bursts\cite{pir1,ant,pir2,ree,wat,ama2}. The best examples of
absorption and emission line are shown in
Figures~\ref{fig:1216},~\ref{fig:0227} and~\ref{fig:0705}.

Even though the reality of the features has been challenged on the
statistical ground, it should not be forgotten that they have been
detected with five different instruments consistently (even though no
single feature has been detected independently in more than one
detector). There are now seven detected narrow features (or system of
features) in emission in the early afterglow\cite{boe} and 2 transient
absorption features during the early prompt emission\cite{ama2}.

\subsection{Emission features}

The detection of bright narrow emission features in the early X-ray
afterglow requires the presence of dense reprocessing material in the
surrounding of the burster\cite{laz3}. The features are detected at
the rest frame frequency of the host galaxy, and cannot therefore be
produced within the fireball, which is still, at that time,
approaching the observer with $\Gamma>10$.

Let us consider as an example case iron lines. The typical line
luminosity is $L_{\rm{Fe}}\sim10^{44}$~erg~s$^{-1}$ and the lines are
detected for a time $t_{\rm{Fe}}\gtrsim10^5$~s. This allows us to
compute the total number of photons in the line as:
\begin{equation}
N_{\rm{Fe}}={{L_{\rm{Fe}}\,t_{\rm{Fe}}}\over{\epsilon_{\rm{Fe}}}}
\sim10^{57}
\end{equation}
which yields a total mass of the reprocessing material of:
\begin{equation}
M_{\rm{tot}}\sim {{m_p\,N_{\rm{Fe}}}\over{A_{\rm{Fe}}\,k}} 
= \frac{2\times10^4}{A_{\rm{Fe},\odot}\,k} \, M_\odot
\end{equation}
where $A_{\rm{Fe}}$ is the iron abundance in number,
$A_{\rm{Fe},\odot}$ is the iron abundance in solar units and $k$ is
the number of line photons that each iron ion emits. It is clear that,
even allowing for a highly iron enriched material, we need
$k\gg10$. An isolated iron ion, taking into account the Auger yield,
can emit up to 12 line photons. The only way to increase the value of
$k$ is by allowing for recombination of free electrons on the iron ion
on a timescale comparable to the burst and early afterglow
duration. Since the recombination timescale is inversely proportional
to the free electron density, we obtain that the density of the
reprocessing material must satisfy:
\begin{equation}
n\gtrsim10^{10} \qquad {\rm cm}^{-3}
\end{equation}

Such a high density poses new problems. It cannot be ahead of the
fireball, since it would slow down the fireball to sub-relativistic
speed in a short time interval, contrary to observations. Two
solutions have been proposed. In the class of Geometry Dominated (GD)
models\cite{laz3,boe2,vie2,laz4,tav}, the reprocessing material is
supposed to be concentrated in an asymmetric toroidal or funnel-like
structure around the GRB explosion site. The most natural progenitor
that could provide such a geometry is the SN remnant of a
Supranova. In the class of Engine Dominated (ED)
models\cite{rees,mes}, the reprocessing material is provided by the
exploding star itself (in the framework of collapsars) and lies behind
the fireball. For this reason the ionising continuum cannot be
provided by burst and/or afterglow photons and a delayed energy
release from the inner engine has to be postulated.

The properties of the two models are discussed elsewhere in these
proceedings\cite{boe} and we therefore discuss here only the problems
facing the ED scenario, given the focus of this contribution in
discussing whether a hypernova origin can be confirmed for all
classical long-duration GRBs.

The first question ED models face is why we see iron lines. Even
though iron lines are the most common features in X-ray spectroscopy,
if the reprocessing material is provided by the exploding star, it
should be rich in nickel rather than in iron\cite{vie2}. This is due to
the fact that in SN explosions $^{56}$Ni is synthesised, which decays
in $^{56}$Co and eventually, after a timescale of $\sim100$~days, into
the stable isotope of $^{56}$Fe. In a simultaneous explosion scenario
the reprocessing material should therefore be nickel rich. There are
two possible solutions to this problem. The first is that there are
areas in the SN explosion parameter space where $^{54}$Fe is
synthesised directly. Such supernovae, however, should have a
particularly dim light curve, lacking the energy input from the decay
of the unstable Ni and Co isotopes. This contradicts the presence of
SN bumps in many afterglow\footnote{It should be clarified here,
however, that in no case to date a simultaneous detection of a SN bump
and of X-ray features has been possible}. Alternatively, it has been
argued that the line may be intrinsically from Ni, but downscattered
to the Fe energy on the walls of the funnel through which the GRB
propagates\cite{macl}. This solution, besides requiring somewhat fine
tuned parameters, faces the problem of efficiency. Each scattering
reduces by a factor of $\sim2$ the number of photons, with the result
that the energy that goes into the line production is at least 100
times larger than the available energy in the GRB explosion\cite{ghi}.

The second problem is why in some afterglow we see iron lines and in
some others we see only lighter elements (Si, S, Ar and Ca). In
the case of GD models, this can be due to a different ionisation
parameter for the reprocessing material\cite{laz4}, as a function of
the different time delay between the SN and the GRB explosion. Short
delays would produce soft X-ray lines, the Fe group ones being
quenched by the Auger effect, while longer delays would produce Fe
lines. In the ED model, there seems to be much less space for changing
parameters in a natural way.

A final problem is represented by the large equivalent width (EW) of
the soft X-ray lines in GRB~011211\cite{ree}. In any reflection model
(optically thin line production is not efficient enough) the EW of the
lines depends on which fraction of the illuminating continuum can reach
the observer. If all the continuum reaches the observer, the EW of
soft X-ray lines has to be small, $\sim150$~eV at most. The only way
to explain the $\sim500$~eV EW detected in 011211\cite{ree} is by
allowing for a pure reflection spectrum, where the ionising continuum
is completely hidden to the observer. In a simultaneous explosion only
a very contrived geometry can provide such a condition, which is
naturally fulfilled in GD models since the line emission reaches the
observer when the continuum has already faded by several orders of
magnitude\cite{laz4}.

\subsection{Absorption features}

The hardest challenge to simultaneous explosion models is however the
presence of transient absorption features in the spectra of at least
one, possibly two, GRBs. These features are detected in the early
phases of the GRB X-ray prompt emission and disappear after tens of
seconds. The best case we have so far is that of GRB~990705\cite{ama2}
in which a deep absorption feature was detected during the first 13
seconds of the prompt emission. The feature, if identified as an Fe
absorption edge, allows a redshift measurement that was subsequently
confirmed by optical spectroscopy of the host galaxy, dramatically
confirming the reality of the feature. The feature was initially
interpreted as an absorption edge from neutral iron\cite{ama2}. Such
an interpretation requires however an extremely large amount of iron
at sub-parsec distances from the burster and was subsequently
abandoned in favour of a broadened resonant feature form highly ionised
iron and cobalt\cite{laz6}. 

The transient nature of the feature is of great importance, since it
allows us to measure the distance of the absorber from the GRB
source\cite{laz5}. The derived distance depends slightly on the
assumed origin of the feature (either edge or resonant line). We make
here the example of the resonant scattering line, which requires a
less dramatic amount of iron group absorbing nuclei. In this case the
absorbing material must be dense enough to allow for a recombination
timescale $t\lesssim10~s$ and the quenching of the absorption is due
not to the complete ionisation of the iron ions but to the heating of
the absorbing medium by Inverse Compton (IC) interactions of the free
electrons with the GRB photons. The IC heating time scale is given
by\cite{laz6}:
\begin{equation}
t_{\rm{IC}}\sim\frac{4\pi\,R^2\,\epsilon}{L\,\sigma_T}
\end{equation}
where $R$ is the distance of the absorber from the burst explosion
site, $\epsilon$ is the typical photon energy and $L$ the burst
luminosity in the keV band. Inserting the relevant numbers for
GRB~990705\cite{ama2} one obtains\cite{laz6}:
\begin{equation}
R\sim2.6\times10^{16} \qquad {\rm cm}
\end{equation}
and, to fulfil the requirement on the recombination time
\begin{equation}
n_e\sim10^{11} \qquad {\rm cm}^{-3}
\end{equation}
Such conditions cannot be satisfied by any stellar wind, however
dense, while they are natural in a moderately clumped young supernova
remnant.

\section{Discussion}

In this paper I have tried to address two questions: is GRB~030329 a
typical GRB so that we can safely claim that all GRBs are associated
to SNe and that the time delay between the two explosions is
negligible? And the second: is there any evidence that is inconsistent
with the above conclusion from independent observations?

The answer to the first question is that indeed GRB~030329 is much
more similar to a classical cosmological GRB that GRB~980425, the
first to be associated to the explosion of a massive star. The
similarity of GRB~030329 with ``classical GRBs'' is however not
complete. First there is evidence that the energy released by
GRB~030329 in gamma-rays is smaller than usual by an order of
magnitude, even though taking into account the energy released by the
inner engine in the form of less relativistic material brings back the
total energy budget to the''standard'' $E=10^{51}$~erg observed in
cosmological GRB explosions\cite{fra,pan}. Secondly, and possibly
related to the delayed energy input of energy, the afterglow
lightcurve is much more complex than any previously observed GRB, and
we showed that this is not due to the more complete sampling, but to
an intrinsic variability that is unprecedented in cosmological GRBs.

An intriguing possibility is that there is a standard inner engine,
possibly a black hole surrounded by a dense hot accretion disk, which
releases a fairly standard energy in the form of a relativistic
jet. The jet has however to open its way into the star and this
creates the diversity in the observed properties of GRBs. Different
relevant properties of the star may be related to its pre-explosion
radius and/or rotation. A compact fast spinning star should offer less
resistance to the jet propagation along its polar axis, giving origin
to a cosmological GRB, in which most of the jet energy can escape
untouched and produce $\gamma$-ray radiation. A more extended or less
rotating star may offer more resistance. In this case a sizable
fraction of the jet energy should be used to open up a funnel in the
star, so that the resulting GRB would look under-energetic. Part, if
not all, this energy may be recycled in a delayed slower fireball
component that would catch up with the relativistic jet and
re-energise it at later times.

This unification picture shall however be taken with care, at least
until a final word is said about the reality of X-ray absorption and
emission features. These cannot be easily incorporated in a single
SN/GRB explosion scenario. Even though none of the claimed features is
incontrovertible in terms of statistical significance, they form a
consistent set of observations all naturally predicted in the two
explosion Supranova scenario\cite{vie}. Instead of having a variable
stellar radius and/or rotation, a unification picture may call for a
variable delay between the two explosions. Some delays are very short,
and produce under-energetic GRBs, since part of the jet energy is
wasted to reach the star surface. Short ($<$~several weeks) delays do
not produce detectable $\gamma$-rays, since the jet is completely
choked inside the optically thick SN remnant\cite{gue}. Longer delays
would instead produce a ``classical GRB'', since the jet has no
baryonic material to cross.

Also this simple scenario faces however some problems, since there are
many ``classical GRBs'' that show sign of red bumps at late times,
usually identified as the emergence of the SN lightcurve on top of the
power-law afterglow decay\cite{blo2,rei,gal2,laz7,blo3}. These SN
explosion should be simultaneous to the GRB explosion. A unification
scenario that would take into account all the observations with their
most probable interpretations would therefore need to take into
account the possibility of both a variability in the progenitor star
properties and of the explosion delay.

\bigskip

{\it I wish to thank S. Covino, F. Frontera, A. K\"onigl,
G. Ghisellini, P. Mazzali, R. Perna, E. Pian, M. J. Rees, E. Rossi,
L. Stella and M. Vietri for the fruitful collaboration and discussions
that led to the development of many of the ideas presented in this
paper.}

\end{document}